\documentclass[]{spie}

\usepackage{amsmath,amsfonts,amssymb}
\usepackage{graphicx}
\usepackage[colorlinks=true, allcolors=blue]{hyperref}

\title{An on-sky investigation into factors limiting the performance of Keck-NIRC2 for conducting infrared high-contrast imaging}

\author[a]{Rachel Bowens-Rubin}
\author[a]{Maissa Salama}
\author[b]{Jayke S. Nguyen}
\author[c]{William Thompson}
\author[a,d]{Philip Hinz}

\affil[a]{University of California Santa Cruz, 1156 High St, Santa Cruz CA, USA}
\affil[b]{University of California San Diego, Astronomy \& Astrophysics Department, 9500 Gilman Dr., San Diego CA 92092, USA}
\affil[c]{University of Victoria}
\affil[d]{University of California Observatories, 1156 High St, Santa Cruz CA, USA}

\authorinfo{Further author information: (Send correspondence to R. Bowens-Rubin)\\R.B.R.: E-mail: rbowens-rubin@mit.edu}

\begin{document} 
\maketitle

\begin{abstract}

The most common instrument used by the exoplanet/brown dwarf direct imaging community at the W.M. Keck Observatory is currently the NIRC2 near-infrared imager. We performed on-sky testing to investigate three effects which may be limiting the performance of NIRC2 when conducting high-contrast imaging observations from $3-5\mu$m
First, we report the measurements of an on-sky test of the throughput of the L/M vector vortex coronagraph. We quantify the throughput and additional background flux penalties, noting that the performance effects of using the vector vortex coronagraph in the Ms-filter are greater than in the Lp-filter. 
Second, we utilize the recently commissioned NIRC2 electronics upgrade to measure the sky variability at sub-second speeds. We find that the background varies at timescales of less than 30\,s, indicating that the electronics upgrade may open an opportunity to improve the sky-background subtraction of future surveys. 
Third, we document the contribution of the image derotator to the spatial non-uniformity in the background flux.
We conclude by giving a set of recommendations of how the Keck-NIRC2 high-contrast imaging community can adapt their observing strategies to improve the sensitivity of future surveys. 

\end{abstract}


\section{INTRODUCTION}
\label{sec:intro}  

The NIRC2 near-infrared imager\footnote{NIRC2 documentation: \url{https://www2.keck.hawaii.edu/inst/nirc2/}} 
is currently the workhorse instrument for the exoplanet/brown dwarf direct imaging community at the W.M. Keck Observatory. NIRC2 is a 1Kx1K Aladdin-3 detector that can operate from 1-5$\mu$m. It is positioned behind the adaptive optics bench on Keck-II telescope.

Direct imaging observations of exoplanets and brown dwarfs are commonly completed with NIRC2 using Lp filter (3.776 $\pm$ 0.700$\mu$m)  \cite{Li2023,Franson2023,Wang2020} 
or the Ms filter (4.670 $\pm$ 0.241$\mu$m) 
\cite{Mawet2019, Llop-Sayson2021,Bowens-Rubin2023,Ren2023,Meshkat2021}. These are the two reddest filters available using NIRC2.\footnote{NIRC2 filter list: (\url{https://www2.keck.hawaii.edu/inst/nirc2/filters.html}}  Using the $3-5\mu$m wavelength range allows for the greatest discovery spaced of cold companions ($<500$K). These wavelengths are also used to distinguish between atmospheric models to measure the disequilibrium carbon chemistry and metallicity \cite{Franson2024}.

In 2023, the readout electronics were upgraded on NIRC2 to decrease the full frame transfer time to save a fits file from 12s to 0.5s, increasing the observing efficiency from 30\% to 75\%.\cite{Alvarez2023}  This upgrade also enabled a high-speed subframing readout. For each science frame, the individual coadds are packaged into a .unp file can be viewed as an image cube.  Typical integration times for coadd frame are between 0.1 - 0.5s with a total image integration time of 30s. The upgrade makes these sub-second timescales available when previously the sampling would only occur on timescales of $>30s$. 
NIRC2 was historically operated with a frame size of of 512 $\times$ 512 pixels (5.090 arcsec $\times$ 5.090 arcsec; pixel scale = 0.009942 $\pm$ 0.00005 arcsec/pixel, \href{https://www2.keck.hawaii.edu/inst/nirc2/genspecs.html}{Keck General Specs}).
However, since the electronics upgrade, observers can now experience little penalty for using the full frame size (1024 x 1024 pixels).  

If the central star is bright and high-contrast is required at small inner working angles ($<0.5$ arcsec) to accomplish the science objective, observers often elect to use the vector vortex coronagraph (VVC) installed at Keck-II to suppress the central starlight \cite{Vortex}.
The first VVC was installed on NIRC2 in March 2015\footnote{ \href{https://www2.keck.hawaii.edu/inst/nirc2/Docs/KAON_1104_Vortex_User_Manual.pdf}{Vortex User Manual}}. Two VVCs have been offered to NIRC2 observers over the past decade, one optimized for K band (\textit{vortexlk}) and another optimized for L/M-band (\textit{vortexlm}).
The centering of the vortex is controlled using the QACTIS  software package \cite{QACITS}. Each QACITS sequence consists of a set of (1) calibration images to acquire an off-axis star PSF and sky images, (2) optimization images to center the star on the vortex and stabilize the tip/tilt in the adaptive optics system, (3) a series of science images.  When the VVCs is used in the optical pathway, observers use the ``fixedhex'' pupil stop which was designed specifically for use with the VVC. To enable angular differential imaging (ADI) post-processing analysis methods with the VVC, NIRC2 observers typically operate Keck in vertical angle mode (with PA $= 4.43^{\circ}$).  

In this proceeding, we investigate three effects which have been suspected to be limiting the performance of high-contrast imaging with NIRC2 in Lp and Ms filters using on-sky testing:  
\begin{itemize}
    \item Section \ref{sec:vortex} is concerned with the  sensitivity impacts introduced by the vector vortex coronagraph. We document the  throughput loss and additional background flux suspected to be introduced by including the coronagraph in the optical path. 
    This effect limits our ability to detect and study colder/smaller/older exoplanets and brown dwarfs at separations $>0.5$ arcsec orbiting bright host stars ($>5$ mag). 
    \item The measurements presented in Section \ref{sec:skybg} are concerned with the speed of the temporal variability in a uniform background due to the changing sky background flux level.   An improved sky background flux correction could  expand the set of colder/smaller/older exoplanets and brown dwarfs that we can study using NIRC2 around both bright and dim stars.  
    \item Section \ref{sec:derotator} documents the added spatial variability caused by the Keck-II image derotator. Understanding and correcting this spatial variability would improve our ability to study cold extended sources such as protoplanetary disks.  
\end{itemize}

\subsection{Background sensitivity expected for NIRC2 with and without coronagraph}

The W.M. Keck Observatory provides an online tool to help observers plan their NIRC2 observations, \href{https://www2.keck.hawaii.edu/inst/nirc2/nirc2_snr_eff.html}{the NIRC2 SNR and Efficiency Calculator}.
This tool provides a theoretically predicted estimate for the SNR of detecting an object of a user-defined magnitude and total exposure time.  The efficiency measurement reported by this tool is no longer accurate as of the 2023 electronics upgrade, but it can still be utilized to assess the expected signal-to-noise ratio. 
The SNR estimate assumes an object is in a background-limited regime of the image and is not affected by another source in the field. Thus, the NIRC2 calculator is not a good predictor of a high-contrast imaging observation at inner working angles tight enough to be in the contrast-limited regime and cannot provide performance estimates when using the vector vortex coronagraph mask. Modified signal-to-noise equations that incorporate the use of the \textit{vortexlm} are presented in the appendix of Bowens-Rubin et al 2023\cite{Bowens-Rubin2023}. 

Observers planning to use the Lp filter in conjunction with the \textit{vortexlm} can utalize the \href{https://wxuan.shinyapps.io/contrast-oracle/}{Vortex Imaging Contrast Oracle (VICO)} \cite{Xuan2018} to predict the expected contrast as a function of separation.  VICO's contrast predictions are based off a training set of 304 targets observed between 2015 to 2018. Because this tool is trained on real data, VICO produces more accurate predictions than the NIRC2 SNR calculator that match on-sky performance. 

For a 1.5 hour total integration time in Lp,
the 5$\sigma$ SNR detection limit predicted by the NIRC2 calculator is 18.2.\footnote{A Strehl=0.85, tint =0.3, coadds=100, Ndithers=180, Nreads=2 were adopted to estimate to sensitivity with the NIRC2 SNR calculator.}  
VICO predicts a background limit of 16.25 when the \textit{vortexlm} is included.\footnote{W1 mag = 5, R mag = 12.35, total integration time = 5400s, and PA rotation = 45deg were adopted to estimate the sensitivity with VICO of a Lp-filter observation with the \textit{vortexlm} in the optical path.} 
In the Ms filter, the SNR calculator predicts that the 5$\sigma$ limit will be at a magnitude of 15.9. 
The published contrast curves plateau at a 5-sigma background limit of $\sim14.2$ when using the Ms filter when the \textit{vortexlm} is included in the optical path (Llop-Sayson et al 2021\cite{Llop-Sayson2021} \& Bowens-Rubin et al 2023\cite{Bowens-Rubin2023}).  

In summary, we find that the prediction of the background limit for non-coronagraphic imaging as compared to background limit as seen during on-sky coronagraphic imaging with the \textit{vortexlm} has a discrepancy of 2.0 mag in the Lp filter and 1.7 mag in the Ms filter. These values correspond to an observing length of $\sim$1.5 hours of total integration, which is a common duration for the length of a NIRC2 observation.   
This discrepancy implies that one or more of the following factors are causing the discrepancy: \textit{(1)} The NIRC2 prediction tools for non-coronagraphic observations are inaccurate, \textit{(2)} the \textit{vortexlm} optic alters the background sensitivity, or \textit{(3)} another instrumentation factor or observing setting used in conjunction with the  \textit{vortexlm} affects the background sensitivity. 
The on-sky testing presented in the following sections aims to trace the accounting of this discrepancy of factors \textit{(2)} and \textit{(3)}.

\section{THROUGHPUT AND THERMAL BACKGROUND EFFECTS OF USING THE VECTOR VORTEX CORONAGRAPH} \label{sec:vortex}

\subsection{Keck Fixedhex pupil stop}
\begin{figure}
    \centering
    \includegraphics[width=0.4\linewidth]{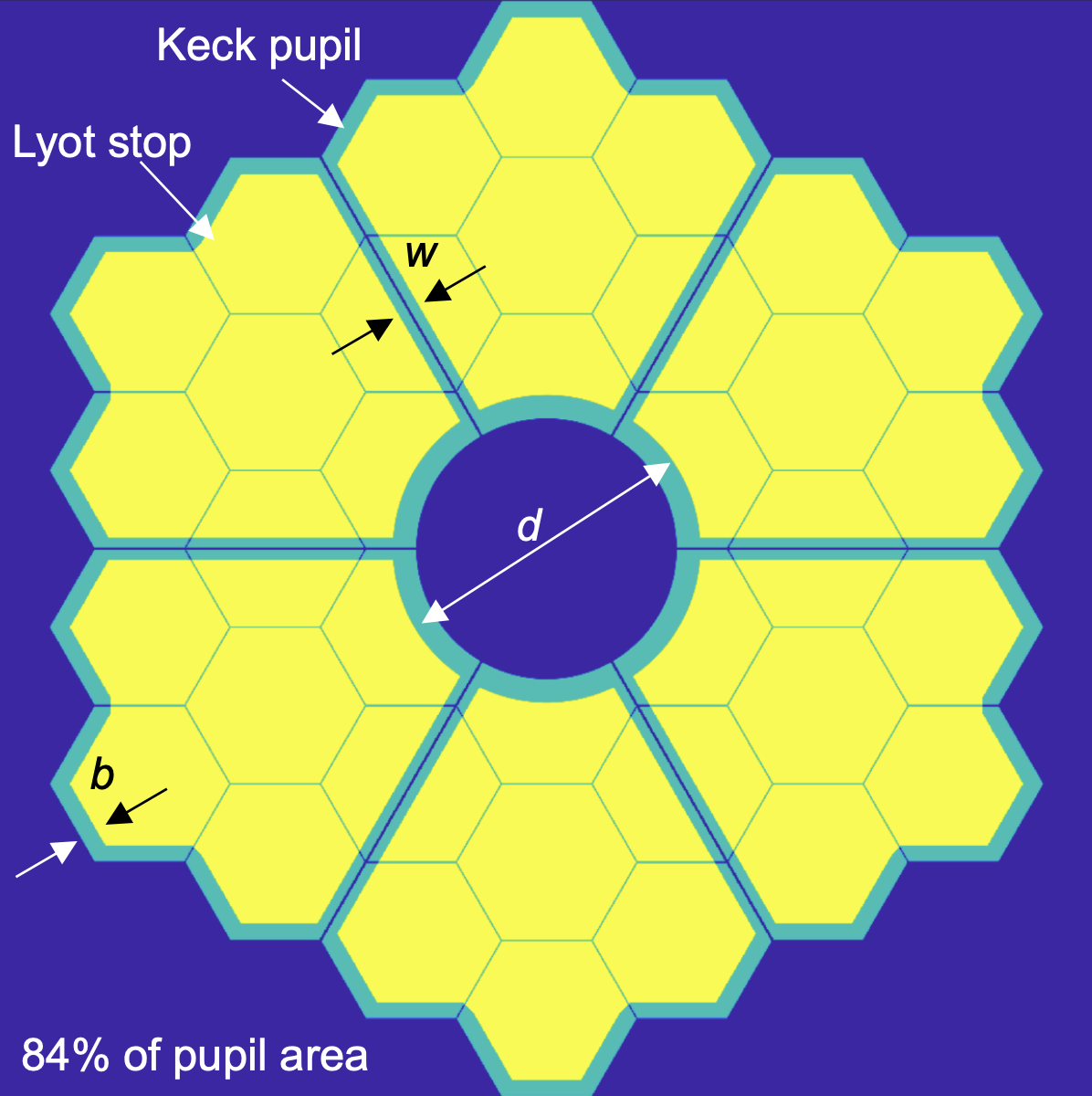}
    \caption{\textbf{Keck fixhex pupil stop.} The fixhex pupil stop is optimized for use with the VVC. It covers 84\% of the pupil area (shown in yellow) as compared to the full size of the Keck pupil (shown in aqua). \textit{Figure Credit: Gary Ruane}}
    \label{fig:fixhex}
\end{figure}

When the \textit{vortexlm} is used in conjunction with NIRC2, observers typically elect to use the fixedhex pupil stop. The fixedhex pupil was designed in 2018 to replace the `incircle' mask for optimization with the vector vortex coronagraph. The new design increased the stop size to let flux pass on the outer edges of the Keck aperture and decreased the stop size near the spiders to exclude thermal flux from the secondary mirror structure spiders. The fixedhex pupil is 84\% the total Keck pupil size (Figure \ref{fig:fixhex}). 

It is not recommended to use an open pupil with the \textit{vortexlm} as this can degrade the overall SNR of the observation. However, observers electing to use the \textit{vortexlm} should remain aware of the additional throughput hit due to the pupil stop choice if they are concerned about the detectablity of their source over the background limit. 

\subsection{Throughput Measurements of the L/M Vector Vortex Coronagraph }

The \textit{vortexlm} is constructed  from a diamond substrate and coated with an antireflective grating.  Diamond absorption varies across the NIRCam operating wavelengths, including an absorption feature that overlaps with the Ms-filter bandpass between 4.5-5$\mu$m  (see Fig. 4 of Webster et al 2015\cite{Webster2015}).  To quantify this variation as it affects the  exoplanet direct imaging community, we performed on-sky throughput testing of the \textit{vortexlm} in Lp and Ms.

Before commissioning, Jolievet et al 2019\cite{Jolivet2019} performed laboratory testing to measure the transmission of the \textit{vortexlm} installed at Keck (referenced in Jolievet 2019 as AGPM-L9r2).  Table 4 of Jolievet et al 2019 reports the transmission of the vortexlm to be 84.7\% $\pm$ 0.9 at L-band (3.575–4.125$\mu$m filter) and $70\pm3$\% at M-band (4.6 $\mu$m laser). 

\begin{figure}
    \centering
    \includegraphics[width=1\linewidth]{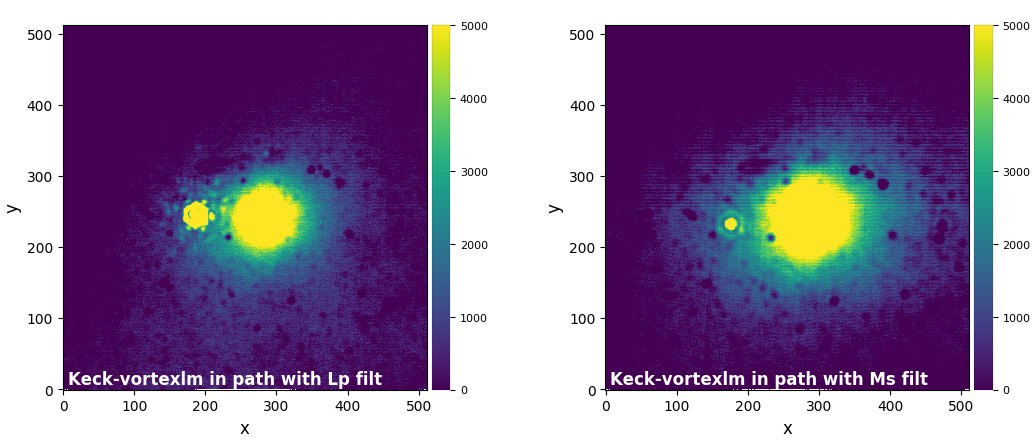}
    \caption{\textbf{Comparison of the PSF vortex-in images with Lp and Ms filters}. The images have been scaled be in units of photons/second, and the sky subtraction was performed by subtracting the median background value. The stellar PSF can be seen on the left and the vortex center glow effect is seen on the right. The vortex center glow is more pronounced in the Ms filter as compared to the Lp filter.    }
    \label{fig:psfvortexin}
\end{figure}

\begin{figure}
    \centering
    \includegraphics[width=0.7\linewidth]{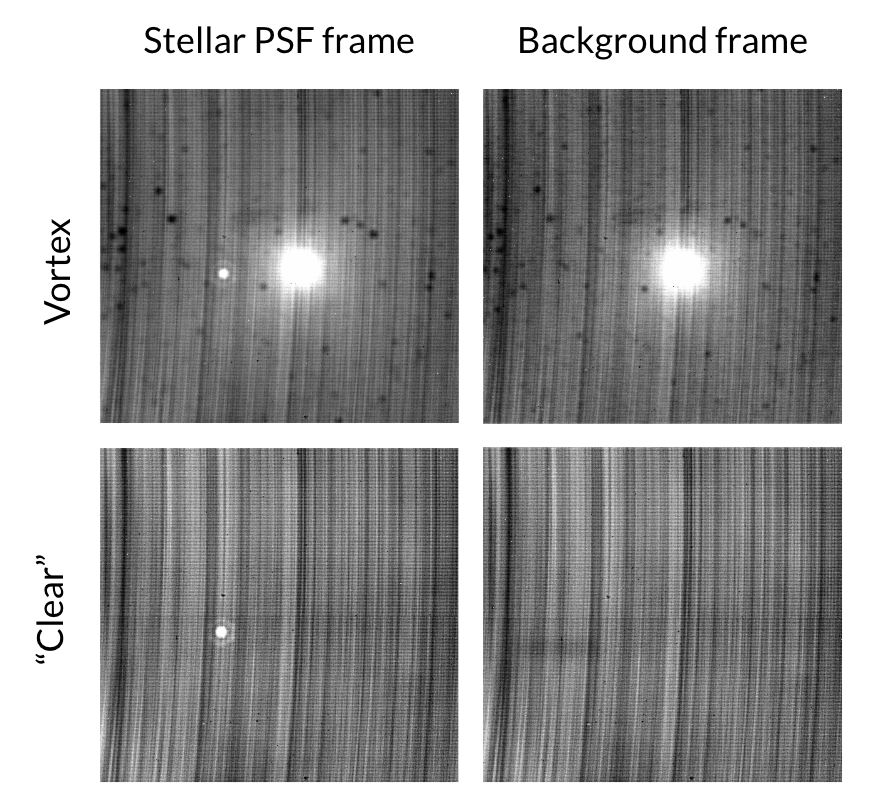}
    \caption{\textbf{Vortex on-sky throughput test.} A star was intentionally miscentered from the vortex to simulate a companion with a separation of 1 arcsec. The four image types were collected to measure the throughput of the \textit{vortexlm} with the Lp and Ms filters.  This figure shows one raw image from each of the four image types from the Ms filter before flat/dark corrections were applied. A dust pattern can be seen to be affecting the spatial  uniformity in the vortex images.}
    \label{fig:throughputtestimgs}
\end{figure}

On August 14 2023 (UT), we measured the throughput of the \textit{vortexlm} on-sky to verify the consistency of the lab throughput measurements.\footnote{These data will soon be publicly available at the Keck Observatory Archive (frames n0197 - n0220): \url{https://koa.ipac.caltech.edu/cgi-bin/KOA/nph-KOAlogin}}  
To complete this testing, we selected a nearby star of sufficient brightness (GJ 4063/Gaia DR3 2110165780975185792; $Hmag=6.5$\cite{2003yCat.2246....0C}) and intentionally miscentered the star from the vortex to simulate a companion with a separation of 1 arcsec (Figure \ref{fig:psfvortexin}). 
A sequence of images were taken in which we moved the vortexlm in and out of the optical path with their corresponding background frames (see Figure \ref{fig:throughputtestimgs}). 
The remaining optics and readout configurations were kept as consistent as possible using the fixedhex pupil, narrow imaging mode, and a 512 x 512 readout.

To complete the on-sky throughput test of the vortexlm, three frames of each were taken in the following sequence for the Lp and Ms filters:
\begin{enumerate}
    \item PSF of star with the \textit{vortexlm} in the optical path 
    \item PSF of star with the \textit{vortexlm} out of the optical path
    \item Background-only (no star) with the \textit{vortexlm} in the optical path
    \item Background-only (no star) with \textit{vortexlm} out of the optical path
\end{enumerate}
This sequence corresponds to 12 frames per filter for a total of 24 frames. Images using the Lp filter were taken with an integration time per coadd of 0.1s with 10 coadds (1s total integration). Images using the Ms filter were taken with an integration time per coadd of 0.2s with 10 coadds (2s total integration). 
Each image sequence was completed in entirety in one filter before starting the image sequence of the next.
The set of three images of the same type were then mean combined to create a representative frame for each image type.

\begin{figure}
    \centering
    \includegraphics[width=1\linewidth]{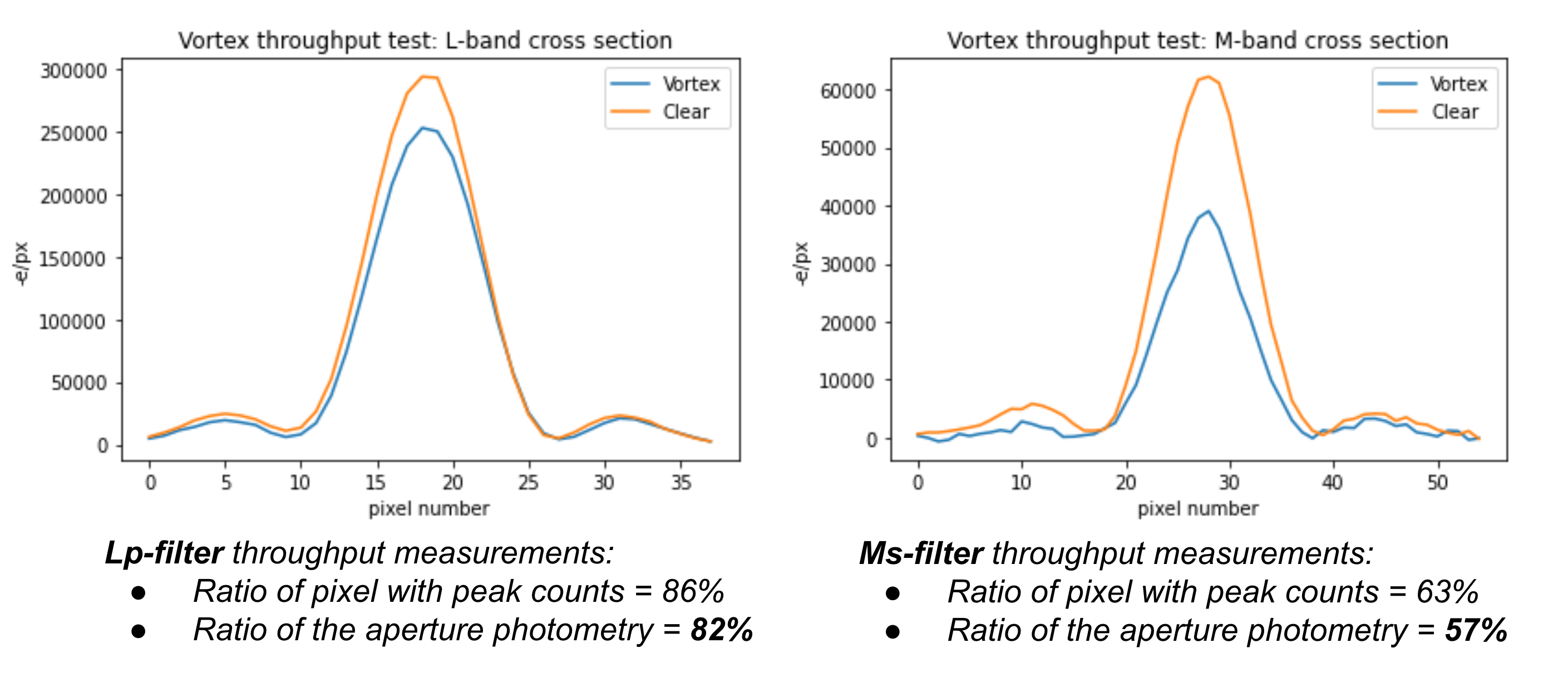}
    \caption{\textbf{Throughput measurement of the vortexlm.}  The comparison of the 1D cross section profiles of the stellar PSF is shown for Lp (\textit{left}) and Ms (\textit{right}). A star was off-centered from the coronagraph by 1 arcsec to act as a proxy for a companion in the background limited regime.  The orange line is the measured flux without a coronagraph, and the blue line is the measured flux with the \textit{vortexlm} in the optical path. For both filters, we see a throughput impact with more impact in Ms.  }
    \label{fig:psfcrosssections}
\end{figure}

To measure the vortex throughput, the background frame was subtracted from its corresponding stellar PSF frame to create a background subtracted PSF frame. The throughput of the star was measured using two methods: \textit{(1)} measuring the ratio between max pixel counts and \textit{(2)} comparing aperture photometry. The aperture photometry was completed using the \texttt{photutils} python package with an aperture radius of 3$\lambda / D$ centered around the stellar PSF. 
Figure \ref{fig:psfcrosssections} shows the cross-section profiles of the stellar PSF with the vortex in and out of the optical path for each filters.   We measured the throughput of the \textit{vortexlm} in Lp to be 86\% using the ratio of the max pixel count and 82\% using aperture photometry, which is consistent with the reported lab results of the AGPM-L9r2 mask ($84.7\% \pm 0.9$). 
We measured the throughput of the \textit{vortexlm} in Ms to be 63\% using the ratio of the max pixel count and 57\% using aperture photometry. This is not consistent with the reported lab measurement for the AGPM-L9r2 mask ($70\% \pm 3\%$), however this result is consistent with the throughput measured of the ensemble of the 9 reported annular groove phase masks ($64.5 \pm 3.5\%$).

\subsection{Spatial non-uniformity \& elevated background counts effects of using the vortexlm}
\begin{figure}
    \centering
\includegraphics[width=0.9\linewidth]{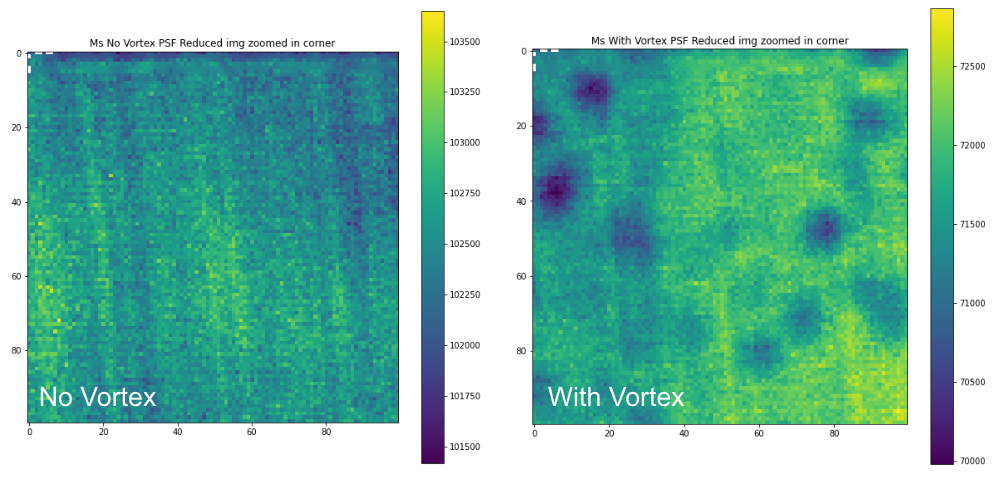}
    \caption{\textbf{Spatial effects on the background due to the vortexlm.} The 100 x 100 pixel zoomed in corner of the stellar PSF frames are shown for the Ms filter.  Spatial non-uniformity in the background is introduced when the \textit{vortexlm} is in the optical path. }
    \label{fig:vortexdust}
\end{figure}

 We compared a 100 $\times$ 100 pixel corner of the stellar PSF frames with and without the coronagraph to examine the effects of including the \textit{vortexlm} in optical path on the background uniformity and flux level (see Figure \ref{fig:vortexdust}).  
  We find evidence for a spatial nonuniformity in the background introduced by the \textit{vortexlm} that looks consistent with dust on the optics. A custom flat field to remove this pattern would be non-trivial to create as it is not standard for observers to take flats with the \textit{vortexlm} in place and the vortex center glow effect causes a non-uniform illumination on the detector. This non-uniformity is most relevant to extended-source science cases, such as measuring the properties of disks. 

The measured mean background counts of the 100 $\times$ 100 pixel section of the frames are reported in Table \ref{tab:vortexthruput}. 
We determined that the ratio of the background counts are elevated above the throughput ratios measured on-sky using stellar PSF. The elevation is  more prominent when using the Ms filter as compared to Lp filter. We hypothesis that this extra source of background flux is added by the \textit{vortexlm} and is due to emission from the optic itself, since M-band has a combination of higher sky background counts (to feed into the emission) and a lower transmission through the \textit{vortexlm} (higher emission coefficient). This elevated background count issue limits the detectability of cold/faint companions in the background limited regime (separation $>0.8''$). 

\begin{figure}
    \centering
    \includegraphics[width=1\linewidth]{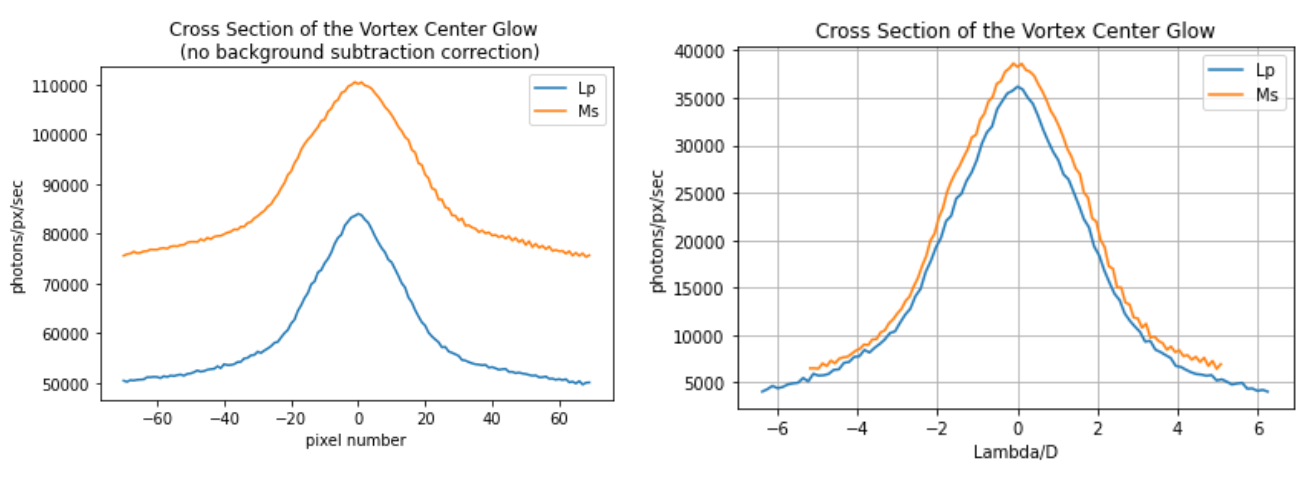}
    \caption{\textbf{Brightness of the Vortex Center Glow.} The 1D cross section along the x-axis of the vortex center glow flux is plotted above in order to compare the center glow in Lp and Ms. \textit{(Left)} The cross sections of the vortex center glow are shown scaled by pixel number before any median background subtraction. \textit{(Right)} A median background subtraction was performed and the cross sections were rescaled by $\lambda/D$. We find the vortex center glow looks similar between Lp and Ms.  The center glow effect is only slightly more present in the Ms filter, differing in max pixel count by 2450 photons/pixel/sec. }
    \label{fig:centerglow}
\end{figure}

It has also been hypothesized that the ``vortex center glow'' effect may cause the elevated background counts issue, however we suggest that this explanation is unlikely. The vortex center glow manifests itself as additional flux on the detector centered in the location of the coronagraph\cite{Shinde2022} (see Figure \ref{fig:centerglow} for the 1D cross-section profile shape).  This glow occurs when off-axis thermal emission from the warm telescope environment is diffracted into the pupil image by the geometry of the vortex coronagraph, similar to the ``reverse'' of the coronagraph's intended purpose to send flux from a bright star to the pupil's edge. The vortex center glow is expected to be most relevant near the vortex center and trail off in intensity by $\sim 5 \lambda/D$. 
We compared the vortex center glow effect in Lp and Ms through a 1D cross section test and find that the intensity of the center glow is similar for the two filters (measured counts are reported in Table \ref{tab:vortexthruput}). The combination of the expectation of the vortex center glow being a localized effect ($< 5 \lambda/D$) along with the similarities displayed between the two filters make the vortex center glow issue an unlikely explanation for the cause of the excess mean background flux.

In a typical high-contrast data reduction, the brightness due to the vortex center glow is removed when subtracting out a sky background frame.  However, while the raw counts can be subtracted, the photon noise from the glow cannot be mitigated through subtraction.  At $2\lambda/D$, the added flux from the vortex center glow is approximately 20000\,phot/px/s. 
Using this number alongside the values reported for the average value for the background with the \textit{vortexlm} in the optical path listed in Table \ref{tab:vortexthruput}, we report that the vortex center glow would reduce the sensitivity by the following for each filter at $2\lambda/D$: Lp = 0.65 mag (83\% of standard background flux) \& Ms = 0.27 mag (28\% of standard background flux).
We find the vortex center glow flux decreases away from the vortex center as expected. This issue most effects the direct imaging detections of companions at small inner working angles ($<0.8$\,arcsec).  

Future instruments with a vector vortex coronagraph that wish to reduce the amount of vortex center glow should include a cold stop  which is not incorporated in the Keck-II AO bench with NIRC2. A cold stop is planned in the optical layouts of Keck-SCALES \cite{Kupke2022} and ELT-METIS \cite{Shinde2022} for this purpose.

\begin{table}[]
\centering
\caption{Vortex On-sky Throughput Testing}
\label{tab:vortexthruput}
\begin{tabular}{|l|c|c|l|}
\hline
 &
  \textbf{Lp} &
  \textbf{Ms} &
  \multicolumn{1}{c|}{\textbf{Notes}} \\ \hline
\begin{tabular}[c]{@{}l@{}}Lab throughput\\ result \cite{Jolivet2019}\end{tabular} &
  84.7 ± 0.9 \% &
  70 ± 3 \% &
  \begin{tabular}[c]{@{}l@{}}$64.5 \pm 3.5\%$ was reported for the ensemble \\ measurement in Ms\end{tabular} \\ \hline
\begin{tabular}[c]{@{}l@{}} On-sky \textit{vortexlm} \\throughput aperture \\photometry ratio \end{tabular} &
  82\% &
  57\% &
  \begin{tabular}[c]{@{}l@{}}On sky testing  measurements are lower than \\ throughput measured in lab\end{tabular} \\ \hline
\begin{tabular}[c]{@{}l@{}}Avg background\\ with \textit{vortexlm}\end{tabular} &
  \begin{tabular}[c]{@{}c@{}}23990 \\ phot/px/sec\end{tabular} &
  \begin{tabular}[c]{@{}c@{}}71730\\ phot/px/sec\end{tabular} &
   \\ \hline
\begin{tabular}[c]{@{}l@{}}Avg background\\ no \textit{vortexlm}\end{tabular} &
  \begin{tabular}[c]{@{}c@{}}28056 \\ phot/px/sec\end{tabular} &
  \begin{tabular}[c]{@{}c@{}}102547\\ phot/px/sec\end{tabular} &
   \\ \hline
\begin{tabular}[c]{@{}l@{}}Measured ratio of  bg\\ counts no/with \textit{vortexlm}\end{tabular} &
  86\% &
  70\% &
  \begin{tabular}[c]{@{}l@{}}Lp background flux slightly elevated from stellar\\  PSF throughput measurement;\\ Ms background flux significantly elevated from stellar \\ PSF throughput measurement but may be \\ consistent with lab test\end{tabular} \\ \hline
\begin{tabular}[c]{@{}l@{}}Peak count\\ added by vortex \\ center glow\end{tabular} &
  \begin{tabular}[c]{@{}c@{}}36182\\ phot/px/sec\end{tabular} &
  \begin{tabular}[c]{@{}c@{}}38632\\ phot/px/sec\end{tabular} &
  \begin{tabular}[c]{@{}l@{}}Peak count added by the vortex center glow is similar\\ between filter, Ms is slightly larger\end{tabular} \\ \hline
\end{tabular}
\end{table}

\section{Sky Background Count Variation at Keck} \label{sec:skybg}

\begin{figure}[]
    \centering
    \includegraphics[width=0.75\linewidth]{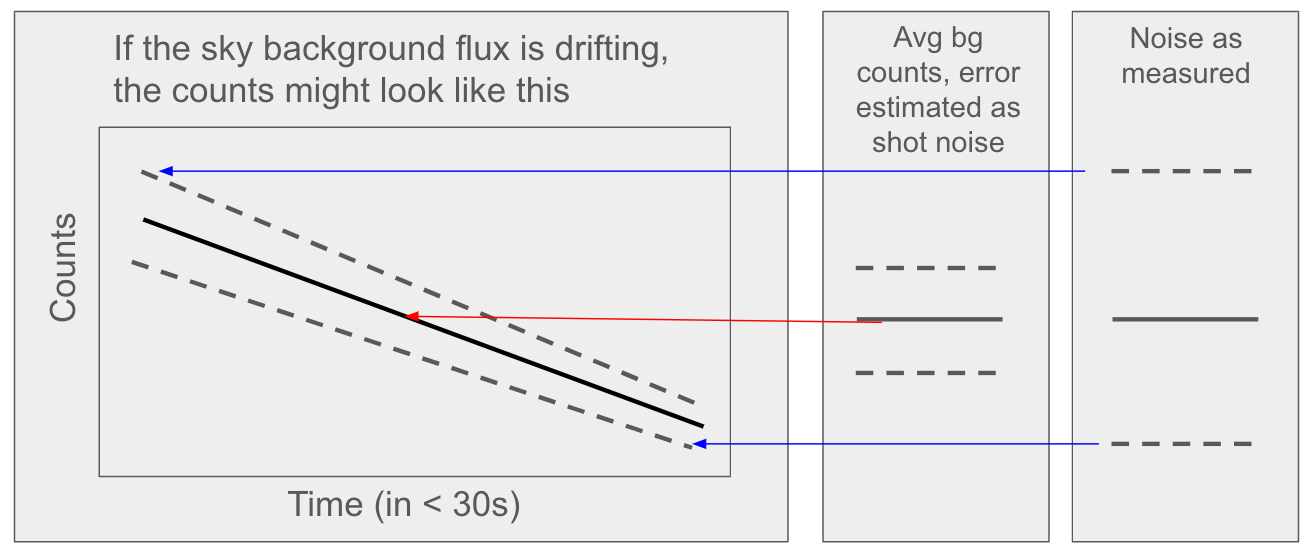}
    \caption{\textbf{Cartoon illustration of how drifts in sky background counts may cause the sky background noise to be under predicted}.  \textit{(Panel 1)} We imagine a drift of sky background counts at a time scale of less than a typical image integration ($<30s$).  The dotted line represents the conceptualized noise level if photon noise is assumed. \textit{(Panel 2 \& 3)} In the final image, we often assume that the mean background counts follows photon noise. However, if there was drifting, the assumption of shot noise may be an underestimate to what is measured.  }
    \label{fig:cartoon}
\end{figure}

During the pre-processing steps of high-contrast imaging analysis, it is typical to remove the flux due to the sky background by performing a subtraction step on each image.  The timescale of each exposure is typically around 30s of total integration.  This background subtraction is often completed using  a uniform median subtraction or using more sophisticated algorithms like PCA sky background subtraction (see Hunziker et al 2018 \cite{Hunziker2018}). In the event that the sky background flux level drifts at speeds faster than the image exposure, the measured noise would be greater than the photon limit estimated for the average background.  Figure \ref{fig:cartoon} shows a cartoon illustration of the phenomenon of the measured error being larger than the predicted error if one were to assign the noise level by adopting the photon-noise limit of the average background flux.

To test the level of sky background drift at sub-30 second timescales, we conducted a set of sky background variation tests over four nights.  A summary of the frames taken to complete the testing is listed in Table \ref{tab:skybgtest}. These frames will be available on the \href{https://koa.ipac.caltech.edu/cgi-bin/KOA/nph-KOAlogin}{Keck Observatory Archive} one year from their listed observation date.    We choose the exposure time of each coadd to have durations of either 0.01s or 0.05s (max speeds achievable with the 64 pixel and 256 subsize respectively).   We used a series of coadd lengths of either 200, 300, or 600.  A coadd length of 300 was used in the second half of testing to try to avoid NIRC2 crashes. Tests were run with both the Lp and Ms filters.  
We analyzed the counts within a 0.2$''$ aperture to resemble the spatial scale used for companion hunting. 

\begin{figure}
    \centering
    \includegraphics[width=1\linewidth]{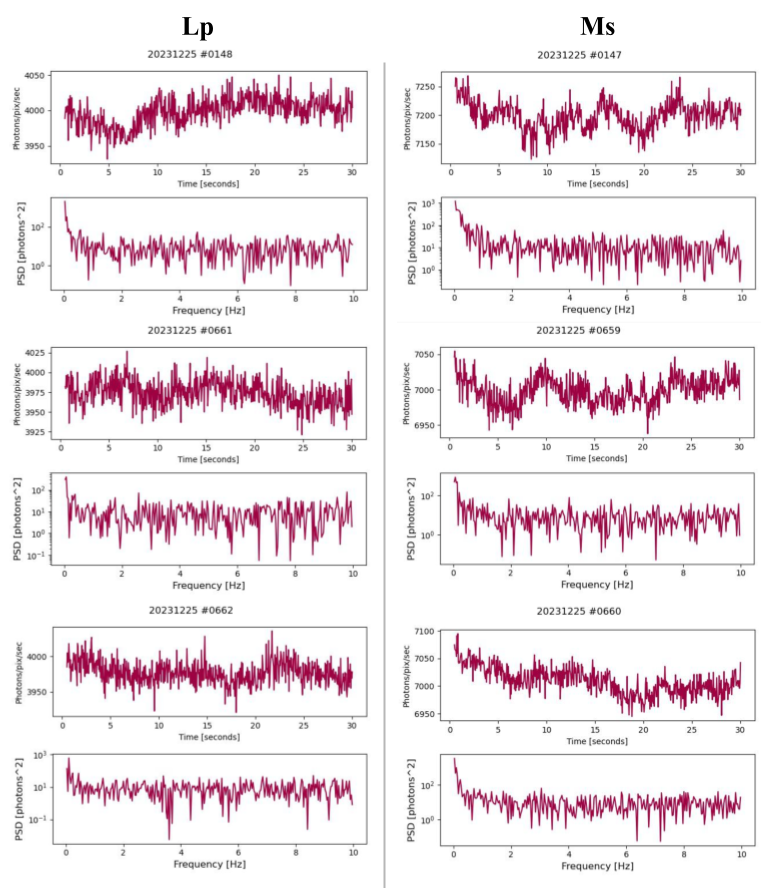}
    \caption{\textbf{Sky Background Variability: Examples of 30s time sequence test from Dec 26 2023 UT.} The background counts measured in each coadd are plotted with their corisponding Power Spectral Density (PSD) signal below for each time series.  }
    \label{fig:skybg20231226}
\end{figure}

Figure \ref{fig:skybg20231226} shows the counts measured as part of the 0.05\,s-speed/30\,s-total test on the evening of Dec 26th 2023. These measurements confirm that flux drifts in the sky background are present at speeds of less than 30\,s and confirm that the 30\,s time sequences are inconsistent with Poisson statistics due to flux drift.  
These background fluctuations are detectable in both the Lp and Ms filters.

\begin{figure}
    \centering
    \includegraphics[width=1\linewidth]{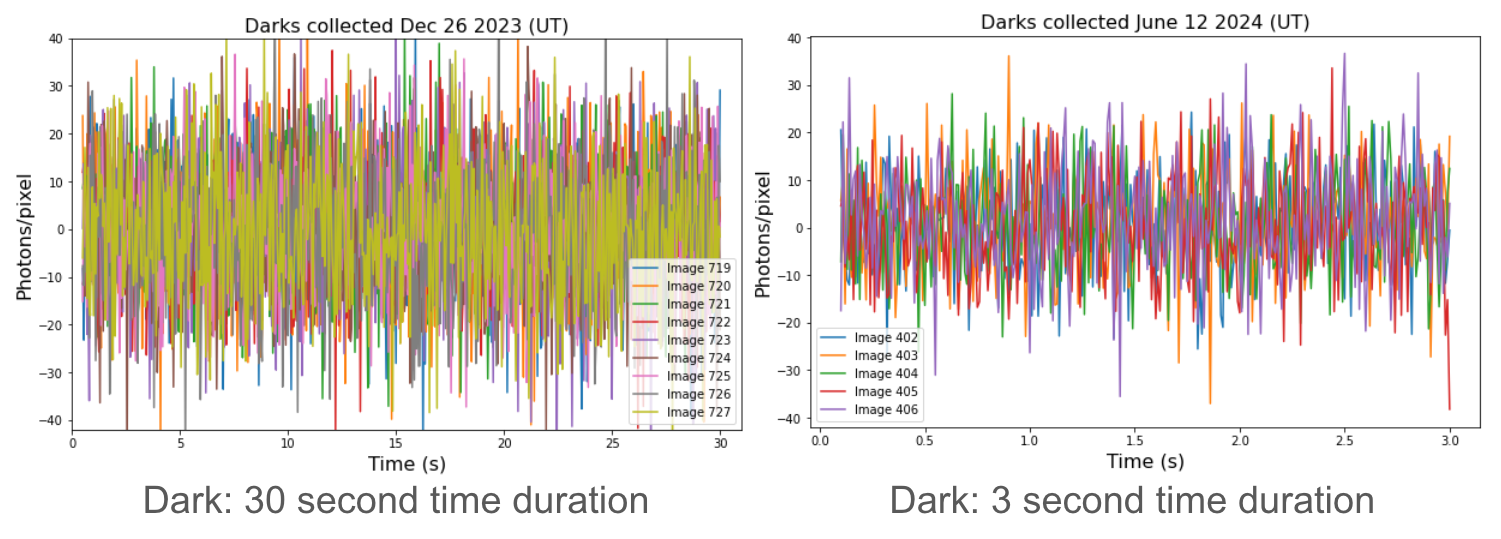}
    \caption{\textbf{Photons per pixel within each coadd of a Dark Frame} We see no drifting in the dark frames of either time duration (tint = 0.05s on left and tint = 0.01s on right). We measured the read noise to be similar within the two sequences   ($\sim12$ photons/px).}
    \label{fig:darkcounts}
\end{figure}

The corresponding dark frames were collected on the four nights. A sampmode of 2 was selected to match typical science images.   Figure \ref{fig:darkcounts} shows the counts measured for each coadd within the dark frames collected on Dec 26 2023 UT and June 12 2024 UT.  
Each individual coadd within the dark frame acts as a bias frame (since sufficient dark current cannot collect in each $<0.05$s coadd). 
We find no notable drifting of counts in time in any set of dark frames. 

To measure read noise, we selected a 20$\times$20 pixel square in a region of the image with no anomalies, summed the counts within the region, and measured the standard deviation of the sums across the frames. 
We measure a read noise of $13.6 \pm 0.5$ photons/pixel using the 10 darks taken in December 26 2023 which has a total time series duration of 30s (coadd = 600, tint = 0.05\,s). We report a read noise of $11.1 \pm 0.4$ photons/pixel using the 5 darks taken in June 12 2024 which have a total time series duration of 3s (coadd = 300, tint = 0.01s). 

These results imply that future observing runs conducted using the newly commissioned NIRC2 electronics could improve their image background subtraction noise if the integration time for each image were reduced. Because there is little read-out noise and read-out time penalties, future high-contrast imaging observations should consider reducing the length of their exposures times to under 30s (optimally $<1$\,s). 

\begin{table}[]
\centering
\caption{Summary of Sky Background Variation Tests}
\label{tab:skybgtest}
\begin{tabular}{ccccccc}
\textbf{UT date} &
  \textbf{\begin{tabular}[c]{@{}c@{}}File \\ num\end{tabular}} &
  \textbf{Filter} &
  \textbf{Coadd} &
  \textbf{\begin{tabular}[c]{@{}c@{}}tint \\ (s)\end{tabular}} &
  \textbf{\begin{tabular}[c]{@{}c@{}}AO Deform. \\ Mirror State\end{tabular}} &
  \textbf{Airmass} \\ \hline
2023-12-26 & 105 & Ms + clear & 200 & 0.05 & closed & 1.33 \\
2023-12-26 & 106 & Lp + clear & 200 & 0.05 & closed & 1.34 \\
2023-12-26 & 147 & Ms + clear & 600 & 0.05 & closed & 1.38 \\
2023-12-26 & 148 & Lp + clear & 600 & 0.05 & closed & 1.4  \\
2023-12-26 & 659 & Ms + clear & 600 & 0.05 & closed & 1.08 \\
2023-12-26 & 660 & Ms + clear & 600 & 0.05 & closed & 1.08 \\
2023-12-26 & 661 & Lp + clear & 600 & 0.05 & closed & 1.09 \\
2023-12-26 & 662 & Lp + clear & 600 & 0.05 & closed & 1.09 \\ \hline
2024-01-24 & 227 & Lp + clear & 600 & 0.05 & closed & 1.05 \\
2024-01-24 & 228 & Lp + clear & 600 & 0.05 & closed & 1.05 \\
2024-01-24 & 229 & Ms + clear & 600 & 0.05 & closed & 1.05 \\
2024-01-24 & 230 & Ms + clear & 600 & 0.05 & closed & 1.05 \\
2024-01-24 & 719 & Lp + clear & 600 & 0.05 & open   & 1.54 \\
2024-01-24 & 720 & Lp + clear & 600 & 0.05 & open   & 1.54 \\
2024-01-24 & 743 & Ms + clear & 600 & 0.05 & open   & 1.54 \\
2024-01-24 & 744 & Ms + clear & 600 & 0.05 & open   & 1.54 \\ \hline
2024-06-12 & 4   & Ms + clear & 300 & 0.01 & open   & 1.46 \\
2024-06-12 & 5   & Ms + clear & 300 & 0.01 & open   & 1.46 \\
2024-06-12 & 6   & Ms + clear & 300 & 0.01 & open   & 1.46 \\
2024-06-12 & 7   & Ms + clear & 300 & 0.05 & open   & 1.46 \\
2024-06-12 & 8   & Ms + clear & 300 & 0.05 & open   & 1.45 \\
2024-06-12 & 9   & Ms + clear & 300 & 0.05 & open   & 1.45 \\
2024-06-12 & 10  & Lp + clear & 300 & 0.01 & open   & 1.45 \\
2024-06-12 & 11  & Lp + clear & 300 & 0.01 & open   & 1.45 \\
2024-06-12 & 12  & Lp + clear & 300 & 0.01 & open   & 1.45 \\
2024-06-12 & 13  & Lp + clear & 300 & 0.05 & open   & 1.44 \\
2024-06-12 & 14  & Lp + clear & 300 & 0.05 & open   & 1.44 \\
2024-06-12 & 15  & Lp + clear & 300 & 0.05 & open   & 1.44 \\ \hline
2024-06-28 & 268 & Lp + clear & 300 & 0.01 & closed & 1.12 \\
2024-06-28 & 269 & Lp + clear & 300 & 0.01 & closed & 1.12 \\
2024-06-28 & 270 & Lp + clear & 300 & 0.01 & closed & 1.12 \\
2024-06-28 & 271 & Ms + clear & 300 & 0.01 & closed & 1.12 \\
2024-06-28 & 272 & Ms + clear & 300 & 0.01 & closed & 1.12 \\
2024-06-28 & 273 & Ms + clear & 300 & 0.01 & closed & 1.12
\end{tabular}
\end{table}

\section{Background contribution of the image derotator} \label{sec:derotator}
During our investigation of the thermal background, a rotating signal matching the rotation of the de-rotator was discovered. The adaptive optics bench on Keck II is located on the nasmyth deck of the telescope, thus requiring an additional optic, a k-mirror, to keep the pupil in the same orientation on the instrument as the target moves throughout the night. This optic, also known as the de-rotator, has the property such that an image field of view rotates by double the drive angle of the de-rotator. Thus resulting in a rotating background signal independent of the atmospheric thermal background and science target. Critically, this background rotates at a different speed than the parallactic angle in ADI mode. Typical background removal techniques using PCA \cite{Hunziker2018} are ineffective at removing this rotating component, since it generally assumes a static background signal.

\begin{figure}
    \centering
    \includegraphics[width=0.85\linewidth]{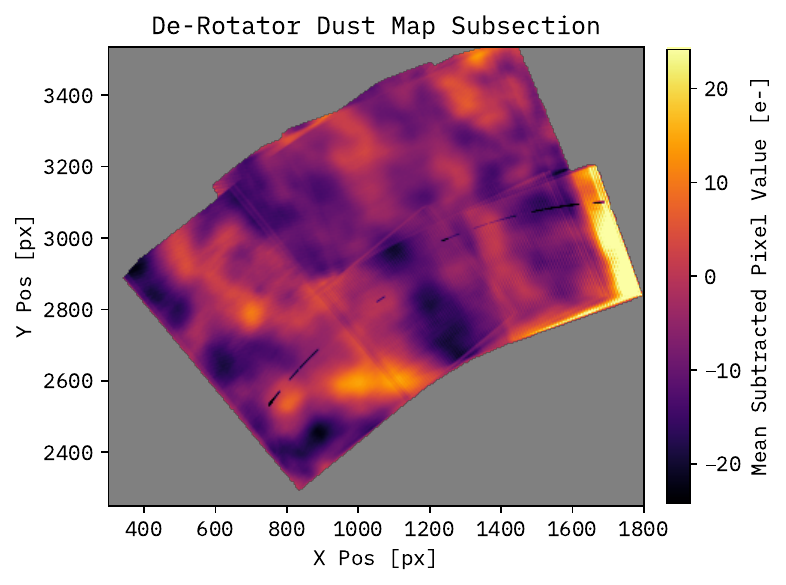}
    \caption{\textbf{The Keck II AO bench de-rotator dust map. }The features present are persistent throughout the observing sequence and are quasi-static, changing slightly between frames. These extended features are not present in shorter wavelength observations.}
    \label{fig:derotator}
\end{figure}

Figure \ref{fig:derotator} shows the de-rotator dust map obtained from median combining a de-rotated stack of science frames taken during an observing sequence on February 20 2024 using NIRC2 on the Keck II telescope. The frames were de-rotated by half of the commanded de-rotator drive angle. The science target was masked out for clarity and these features are found to be consistent with the de-rotator reference frame.

We speculate that the origin of these features come from dust on the de-rotator itself. Although the k-mirror is in an enclosure, there is air flowing through the AO bench. The last cleaning of the de-rotator on the Keck II AO bench occurred in April 2017 therefore, the optics are expected to have collected some amount of dust. Another source of this emission may be optical imperfections such as scratches or dents on the k-mirror itself.

Further work involving the origin and impact of the de-rotator dust map on mid-IR observations are ongoing.

\section{CONCLUSIONS}

Based on the on-sky testing documented in this proceeding, we make the following recommendations for observers and instrument staff seeking to improve the high-contrast imaging performance of Keck/NIRC2:

\begin{itemize}

\item The use of the vortex coronagraph should be carefully evaluated when planning observations, as it reduces throughput (reductions seen $>15$\%) and increases the flux present in the image background. The throughput and increased background effects are significantly more pronounced in the Ms filter than in the Lp filter (see Table \ref{tab:vortexthruput}).

\item A ``vortex center glow'' effect is present which is caused by off-axis thermal emission being diffracted into the NIRC2 image by the \textit{vortexlm}.  The flux level of the vortex center glow is similar between the Lp and Ms filters.

\item The spatial non-uniformity of the \textit{vortexlm} and image rotator may be mitigated through regular optics cleaning to remove dust. For science cases limited by this non-uniformity, a dedicated flat-fielding procedure may improve sensitivity. 

\item Quick-readout mode tests with the upgraded NIRC2 electronics show that background counts in Lp and Ms drift on timescales shorter than 30s. Using exposure times shorter than 30s may improve background subtraction for Keck instruments in these bands.
\item Preliminary evidence suggests the image derotator introduces a quasi-static background pattern, with or without the coronagraph. No standard calibration method currently exists, but careful procedures could help reduce these effects. More frequent cleaning of the optic may also improve performance. 

\item We recommend that observatory staff measure background performance before and after the next cleaning of both the rotator and the vortex coronagraph to quantify the benefits of regular cleaning.
\end{itemize}

By addressing these challenges now with Keck NIRC2, we have the opportunity to not only improving our current high-contrast imaging capabilities, but lay the critical groundwork that will guide the design and operation of the high-contrast imaging instruments at ELT-class facilities.

\acknowledgments 
The authors would like to thank the contributions of the Keck staff and collaborators who provided their valueable technical expertise and opinions: 
Carlos Alvarez, Olivier Absil, Antonin Bouchez, Steph Sallum, Andy Skemer, Brittnay Miles, and Jarron Leisenring.

\bibliography{report} 

\begin{thebibliography}{10}

\bibitem{Li2023}
{Li}, Y., {Brandt}, T.~D., {Brandt}, G.~M., {An}, Q., {Franson}, K., {Dupuy}, T.~J., {Chen}, M., {Bowens-Rubin}, R., {Lewis}, B.~L., {Bowler}, B.~P., {Gibbs}, A., {Kiman}, R., {Faherty}, J., {Currie}, T., {Jensen-Clem}, R., {Zhang}, H., {Contreras-Martinez}, E., {Fitzgerald}, M.~P., {Mazin}, B.~A., and {Millar-Blanchaer}, M., ``{Surveying nearby brown dwarfs with HGCA: direct imaging discovery of a faint, high-mass brown dwarf orbiting HD 176535 A},'' {\em mnras}~{\bf 522},  5622--5637 (July 2023).

\bibitem{Franson2023}
{Franson}, K., {Bowler}, B.~P., {Zhou}, Y., {Pearce}, T.~D., {Bardalez Gagliuffi}, D.~C., {Biddle}, L., {Brandt}, T.~D., {Crepp}, J.~R., {Dupuy}, T.~J., {Faherty}, J., {Jensen-Clem}, R., {Morgan}, M., {Sanghi}, A., {Theissen}, C.~A., {Tran}, Q.~H., and {Wolf}, T.~A., ``{Astrometric Accelerations as Dynamical Beacons: A Giant Planet Imaged Inside the Debris Disk of the Young Star AF Lep},'' {\em arXiv e-prints} ,  arXiv:2302.05420 (Feb. 2023).

\bibitem{Wang2020}
{Wang}, J.~J., {Ginzburg}, S., {Ren}, B., {Wallack}, N., {Gao}, P., {Mawet}, D., {Bond}, C.~Z., {Cetre}, S., {Wizinowich}, P., {De Rosa}, R.~J., {Ruane}, G., {Liu}, M.~C., {Absil}, O., {Alvarez}, C., {Baranec}, C., {Choquet}, {\'E}., {Chun}, M., {Defr{\`e}re}, D., {Delorme}, J.-R., {Duch{\^e}ne}, G., {Forsberg}, P., {Ghez}, A., {Guyon}, O., {Hall}, D. N.~B., {Huby}, E., {Jolivet}, A., {Jensen-Clem}, R., {Jovanovic}, N., {Karlsson}, M., {Lilley}, S., {Matthews}, K., {M{\'e}nard}, F., {Meshkat}, T., {Millar-Blanchaer}, M., {Ngo}, H., {Orban de Xivry}, G., {Pinte}, C., {Ragland}, S., {Serabyn}, E., {Catal{\'a}n}, E.~V., {Wang}, J., {Wetherell}, E., {Williams}, J.~P., {Ygouf}, M., and {Zuckerman}, B., ``{Keck/NIRC2 L'-band Imaging of Jovian-mass Accreting Protoplanets around PDS 70},'' {\em aj}~{\bf 159},  263 (June 2020).

\bibitem{Mawet2019}
{Mawet}, D., {Hirsch}, L., {Lee}, E.~J., {Ruffio}, J.-B., {Bottom}, M., {Fulton}, B.~J., {Absil}, O., {Beichman}, C., {Bowler}, B., {Bryan}, M., {Choquet}, E., {Ciardi}, D., {Christiaens}, V., {Defr{\`e}re}, D., {Gomez Gonzalez}, C.~A., {Howard}, A.~W., {Huby}, E., {Isaacson}, H., {Jensen-Clem}, R., {Kosiarek}, M., {Marcy}, G., {Meshkat}, T., {Petigura}, E., {Reggiani}, M., {Ruane}, G., {Serabyn}, E., {Sinukoff}, E., {Wang}, J., {Weiss}, L., and {Ygouf}, M., ``{Deep Exploration of Eps Eridani with Keck Ms-band Vortex Coronagraphy and Radial Velocities: Mass and Orbital Parameters of the Giant Exoplanet},'' {\em aj}~{\bf 157},  33 (Jan. 2019).

\bibitem{Llop-Sayson2021}
{Llop-Sayson}, J., {Wang}, J.~J., {Ruffio}, J.-B., {Mawet}, D., {Blunt}, S., {Absil}, O., {Bond}, C., {Brinkman}, C., {Bowler}, B.~P., {Bottom}, M., {Chontos}, A., {Dalba}, P.~A., {Fulton}, B.~J., {Giacalone}, S., {Hill}, M., {Hirsch}, L.~A., {Howard}, A.~W., {Isaacson}, H., {Karlsson}, M., {Lubin}, J., {Madurowicz}, A., {Matthews}, K., {Morris}, E., {Perrin}, M., {Ren}, B., {Rice}, M., {Rosenthal}, L.~J., {Ruane}, G., {Rubenzahl}, R., {Sun}, H., {Wallack}, N., {Xuan}, J.~W., and {Ygouf}, M., ``{Constraining the Orbit and Mass of epsilon Eridani b with Radial Velocities, Hipparcos IAD-Gaia DR2 Astrometry, and Multiepoch Vortex Coronagraphy Upper Limits},'' {\em aj}~{\bf 162},  181 (Nov. 2021).

\bibitem{Bowens-Rubin2023}
{Bowens-Rubin}, R., {Akana Murphy}, J.~M., {Hinz}, P.~M., {Limbach}, M.~A., {Seifahrt}, A., {Kiman}, R., {Salama}, M., {Mukherjee}, S., {Brady}, M., {Carter}, A.~L., {Jensen-Clem}, R., {van Kooten}, M. A.~M., {Isaacson}, H., {Kosiarek}, M., {Bean}, J.~L., {Kasper}, D., {Luque}, R., {Stef{\'a}nsson}, G., and {St{\"u}rmer}, J., ``{A Wolf 359 in Sheep's Clothing: Hunting for Substellar Companions in the Fifth-closest System Using Combined High-contrast Imaging and Radial Velocity Analysis},'' {\em aj}~{\bf 166},  260 (Dec. 2023).

\bibitem{Ren2023}
{Ren}, B.~B., {Wallack}, N.~L., {Hurt}, S.~A., {Mawet}, D., {Carter}, A.~L., {Echeverri}, D., {Llop-Sayson}, J., {Meshkat}, T., {Oppenheimer}, R., {Aguilar}, J., {Cady}, E., {Choquet}, {\'E}., {Ruane}, G., {Vasisht}, G., and {Ygouf}, M., ``{Planet search with the Keck/NIRC2 vortex coronagraph in the M$_{s}$ band for Vega},'' {\em aap}~{\bf 670},  A162 (Feb. 2023).

\bibitem{Meshkat2021}
{Meshkat}, T., {Gao}, P., {Lee}, E.~J., {Mawet}, D., {Choquet}, E., {Ygouf}, M., {Patel}, R., {Ruane}, G., {Wang}, J., {Wallack}, N., {Absil}, O., and {Beichman}, C., ``{Characterization of HD 206893 B from Near- to Thermal-infrared},'' {\em apj}~{\bf 917},  62 (Aug. 2021).

\bibitem{Franson2024}
{Franson}, K., {Balmer}, W.~O., {Bowler}, B.~P., {Pueyo}, L., {Zhou}, Y., {Rickman}, E., {Zhang}, Z., {Mukherjee}, S., {Pearce}, T.~D., {Bardalez Gagliuffi}, D.~C., {Biddle}, L.~I., {Brandt}, T.~D., {Bowens-Rubin}, R., {Crepp}, J.~R., {Davidson}, James~W., J., {Faherty}, J., {Ginski}, C., {Horch}, E.~P., {Morgan}, M., {Morley}, C.~V., {Perrin}, M.~D., {Sanghi}, A., {Salama}, M., {Theissen}, C.~A., {Tran}, Q.~H., and {Wolf}, T.~N., ``{JWST/NIRCam 4-5 $\mu$m Imaging of the Giant Planet AF Lep b},'' {\em arXiv e-prints} ,  arXiv:2406.09528 (June 2024).

\bibitem{Alvarez2023}
{Alvarez}, C., {Kassis}, M., {Greffe}, T., {Smith}, R., {Baril}, M., {Campbell}, R., {Gomez}, P., {Kirby}, E., {Mawet}, D., {Roberts}, M., {Weber}, R., {Hale}, D., {Cavalieri}, D., {Holewczynski}, J., {Smous}, J., {Kwok}, S., {Chan}, D., and {Dahler}, M., ``{Upgrades to W. M. Keck observatory detector systems},'' {\em Astronomische Nachrichten}~{\bf 344},  e20230062 (Oct. 2023).

\bibitem{Vortex}
{Serabyn}, E., {Huby}, E., {Matthews}, K., {Mawet}, D., {Absil}, O., {Femenia}, B., {Wizinowich}, P., {Karlsson}, M., {Bottom}, M., {Campbell}, R., {Carlomagno}, B., {Defr{\`e}re}, D., {Delacroix}, C., {Forsberg}, P., {Gomez Gonzalez}, C., {Habraken}, S., {Jolivet}, A., {Liewer}, K., {Lilley}, S., {Piron}, P., {Reggiani}, M., {Surdej}, J., {Tran}, H., {Vargas Catal{\'a}n}, E., and {Wertz}, O., ``{The W. M. Keck Observatory Infrared Vortex Coronagraph and a First Image of HIP 79124 B},'' {\em aj}~{\bf 153},  43 (Jan. 2017).

\bibitem{QACITS}
{Huby}, E., {Bottom}, M., {Femenia}, B., {Ngo}, H., {Mawet}, D., {Serabyn}, E., and {Absil}, O., ``{On-sky performance of the QACITS pointing control technique with the Keck/NIRC2 vortex coronagraph},'' {\em aap}~{\bf 600},  A46 (Apr. 2017).

\bibitem{Xuan2018}
{Xuan}, W.~J., {Mawet}, D., {Ngo}, H., {Ruane}, G., {Bailey}, V.~P., {Choquet}, {\'E}., {Absil}, O., {Alvarez}, C., {Bryan}, M., {Cook}, T., {Femen{\'\i}a Castell{\'a}}, B., {Gomez Gonzalez}, C., {Huby}, E., {Knutson}, H.~A., {Matthews}, K., {Ragland}, S., {Serabyn}, E., and {Zawol}, Z., ``{Characterizing the Performance of the NIRC2 Vortex Coronagraph at W. M. Keck Observatory},'' {\em aj}~{\bf 156},  156 (Oct. 2018).

\bibitem{Webster2015}
{Webster}, S., {Chen}, Y., {Turri}, G., {Bennett}, A., {Wickham}, B., and {Bass}, M., ``{Intrinsic and extrinsic absorption of chemical vapor deposition single-crystal diamond from the middle ultraviolet to the far infrared},'' {\em Journal of the Optical Society of America B Optical Physics}~{\bf 32},  479 (Mar. 2015).

\bibitem{Jolivet2019}
{Jolivet}, A., {Orban de Xivry}, G., {Huby}, E., {Piron}, P., {Catalan}, E.~V., {Habraken}, S., {Surdej}, J., {Karlsson}, M., and {Absil}, O., ``{L- and M-band annular groove phase mask in lab performance assessment on the vortex optical demonstrator for coronagraphic applications},'' {\em Journal of Astronomical Telescopes, Instruments, and Systems}~{\bf 5},  025001 (Apr. 2019).

\bibitem{2003yCat.2246....0C}
{Cutri}, R.~M., {Skrutskie}, M.~F., {van Dyk}, S., {Beichman}, C.~A., {Carpenter}, J.~M., {Chester}, T., {Cambresy}, L., {Evans}, T., {Fowler}, J., {Gizis}, J., {Howard}, E., {Huchra}, J., {Jarrett}, T., {Kopan}, E.~L., {Kirkpatrick}, J.~D., {Light}, R.~M., {Marsh}, K.~A., {McCallon}, H., {Schneider}, S., {Stiening}, R., {Sykes}, M., {Weinberg}, M., {Wheaton}, W.~A., {Wheelock}, S., and {Zacarias}, N., ``{VizieR Online Data Catalog: 2MASS All-Sky Catalog of Point Sources (Cutri+ 2003)}.'' VizieR On-line Data Catalog: II/246. Originally published in: University of Massachusetts and Infrared Processing and Analysis Center, (IPAC/California Institute of Technology) (2003) (June 2003).

\bibitem{Shinde2022}
{Shinde}, M., {Delacroix}, C., {Orban de Xivry}, G., {Absil}, O., and {van Boekel}, R., ``{Modeling the vortex center glow in the ELT/METIS vortex coronagraph},'' in [{\em Modeling, Systems Engineering, and Project Management for Astronomy X}{\nolinebreak\hspace{0.1em}]},  {Angeli}, G.~Z. and {Dierickx}, P., eds., {\em Society of Photo-Optical Instrumentation Engineers (SPIE) Conference Series} {\bf 12187},  121870E (Aug. 2022).

\bibitem{Kupke2022}
{Kupke}, R., {Stelter}, R.~D., {Hasan}, A., {Surya}, A., {Kain}, I., {Briesemeister}, Z., {Li}, J., {Hinz}, P., {Skemer}, A., {Gerard}, B., {Dillon}, D., and {Ratliff}, C., ``{SCALES on Keck: optical design},'' in [{\em Ground-based and Airborne Instrumentation for Astronomy IX}{\nolinebreak\hspace{0.1em}]},  {Evans}, C.~J., {Bryant}, J.~J., and {Motohara}, K., eds., {\em Society of Photo-Optical Instrumentation Engineers (SPIE) Conference Series} {\bf 12184},  121844A (Aug. 2022).

\bibitem{Hunziker2018}
{Hunziker}, S., {Quanz}, S.~P., {Amara}, A., and {Meyer}, M.~R., ``{PCA-based approach for subtracting thermal background emission in high-contrast imaging data},'' {\em aap}~{\bf 611},  A23 (Mar. 2018).

\end{thebibliography}
\bibliographystyle{spiebib} 

\end{document}